\UseRawInputEncoding
\documentclass[twocolumn,aps,prb,groupedaddress,amsfonts,amssymb,amsmath]{revtex4}
\usepackage{graphicx}

\begin{document}

\draft

\title{Photogalvanic effect induced charge and spin photocurrent in group-V monolayer systems}

%Photogalvanic effect in group-V monolayers

\author{Li-Wen Zhang$^{1}$, Ya-Qing Yang$^{2,4}$, Jun Chen$^{3,4}$, Lei Zhang$^{2,4,\dagger}$}
\address{$^1$School of Physics and Information Engineering, Shanxi Normal University, Taiyuan 030031, China\\
$^2$State Key Laboratory of Quantum Optics and Quantum Optics Devices, Institute of Laser Spectroscopy, Shanxi University, Taiyuan 030006, China\\
$^3$State Key Laboratory of Quantum Optics and Quantum Optics Devices, Institute of Theoretical Physics, Shanxi University, Taiyuan 030006, China\\
$^4$Collaborative Innovation Center of Extreme Optics, Shanxi University, Taiyuan 030006, China}
%Corresponding author. E-mail: $^\dagger$zhanglei@sxu.edu.cn

%\date{\today}

\begin{abstract}
Photogalvanic effect (PGE) occurs in materials with non-centrosymmetric structures when irradiated by linearly or circularly polarized light. Here, using non-equilibrium Green's function combined with density functional theory (NEGF-DFT), we investigated the linear photogalvanic effect (LPGE) in monolayers of group-V elements (As, Sb, and Bi) by first-principles calculations. First, by designing a two-probe structure based on the group-V elements, we found a giant anisotropy photoresponse of As between the armchair and zigzag directions. Then, we analyzed Sb and Bi's charge and spin photocurrent characteristics when considering the spin-orbit coupling (SOC) effect. It is found that when the polarization direction of linearly polarized light is parallel or perpendicular to the transport direction ($\theta$ = $0^ \circ$ or $90^ \circ$), the spin up and spin down photoresponse in the armchair direction has the same magnitude and direction, leading to the generation of net charge current. However, in the zigzag direction, the spin up and spin down photoresponse have the same magnitude with opposite directions, leading to the generation of pure spin current. Furthermore, it is understood by analyzing the bulk spin photovoltaic (BSPV) coefficient from the symmetry point of view. Finally, we found that the net charge current generated in the armchair direction and the pure spin current generated in the zigzag direction can be further tuned with the increase of the material's buckling height $|h|$. Our results highlight that these group-V monolayers are promising candidates for novel functional materials, which will provide a broad prospect for the realization of ultrathin ferroelectric devices in optoelectronics due to their spontaneous polarization characteristics and high Curie temperature.
\end{abstract}

%\textbf{Keywords} group-V monolayers, Photogalvanic effect, pure spin current

\maketitle
\section{Introduction}
Ferroelectric materials, with a spontaneous polarization, have attracted significant attention both experimentally and theoretically in recent years due to their potential applications, including field effect transistors, non-volatile memory devices, solar cells and sensors, etc\cite{Scott2007, Dawber2005, martin2017, jiang2016, tao2016}. However, it is well known that the spontaneous polarization in ferroelectric materials can lead to charge accumulation on the surfaces of the materials. For three-dimensional (3D) ferroelectric materials, such as $BiFeO_3$, $BaTiO_3$, etc., which tend to lose their ferroelectricity when the thickness is less than a critical value due to the effect of the unscreened depolarizing electrostatic field\cite{ahn2004, fong2004, junquera2003}. Therefore, it cannot satisfy the technological demand of ongoing device miniaturization. Since then, exploring two-dimensional (2D) ferroelectric materials has been essential to overcome the problem\cite{kang2019, liu2018}.

As an element of group-V, black phosphorus has attracted extensive attention in recent years due to its tunable direct band gap in a wide range (from visible to infrared), high carrier mobility, and other characteristics\cite{Li2021,Qiao2014, Gong2014, zhang2018, Liu2014, zhang2019, Tu2014, li2014, buscema2014, Avsar2017}. It has been applied in the preparation of field effect transistors, spin valves, memory, etc\cite{zhang2018, Liu2014, zhang2019, Tu2014, li2014, buscema2014, Avsar2017}. The structure of monolayer black phosphorus preserves spatial inversion symmetry, hence forbidding ferroelectric (FE). The FE may be induced by extrinsic means, e.g., via an external electric field or by substituting different elements for P to break the centrosymmetry of the structure. Recently, Xiao et al. revealed that 2D elemental group-V (As, Sb, Bi) monolayers with spontaneous polarization due to the lattice distortion with atomic layer buckling and has quite sizable values, which are comparable to or even larger than some 2D monolayer compounds\cite{xiao2018, wan2017}. Their Curie temperatures can be higher than room temperature, making them promising candidates for ultrathin ferroelectric devices\cite{xiao2018}. In addition, the puckered lattice structure of group-V monolayers, such as Sb and Bi, have already been experimentally demonstrated\cite{bianchi2012, lu2015, kokubo2015}. All of these provide a strong guarantee for the theoretical research on the properties of group-V monolayers. It is well known that the photogalvanic effect (PGE) occurs in materials without spatial inversion symmetry under the illumination of polarized light\cite{Liu2022, Qian2022, chen2018, xie2015, ganichev2002-2, li2017, zhao2017, jiang2011, guan2017, belinicher1980}. Because of the lack of centrosymmetry of the group-V (As, Sb, Bi) monolayer with the puckered lattice structures, the distribution of the photo-excited electrons in the conduction bands is imbalanced, which leads to a persistent photocurrent without the need to apply an external bias voltage or a temperature gradient. All these clues motivate us to investigate the PGE of the two-probe devices based on the group-V monolayer. Moreover, the spin-orbit coupling (SOC) are noticeable for Sb and Bi atoms. And there are many theoretical progresses on the generation of pure spin current by the PGE recently\cite{Tao2020,Shu2022,Ishizuka2022,Fei2021}. Thus, the following questions naturally arise: Can the photo-induced pure spin current be generated and further tuned?

In this work, we answer the question by investigating the linear PGE (LPGE) of two-probe devices based on the group-V monolayer. Due to the lack of centrosymmetry, finite photocurrent can be obtained without needing any external electric field in the system. By analyzing charge and spin photocurrent characteristics of Sb and Bi when considering the spin-orbit coupling (SOC) effect, we found that net charge current and pure spin current can be generated in the armchair and zigzag direction, respectively, when the polarization direction of linearly polarized light is parallel or perpendicular to the transport direction ($\theta$ = $0^ \circ$ or $90^ \circ$). This is due to the system posses mirror symmetry $M^x$. In addition, by calculating the photoresponse of different buckling heights, we also found that the net charge current generated in the armchair direction and the pure spin current generated in the zigzag direction can be further tuned with the increase of the material's buckling height $|h|$. The above conclusions provide a theoretical basis for the study of group-V monolayer ferroelectric materials in optoelectronic devices.

%The rest of the paper is organized as follows. In Section 2, we present the computational details and theoretical formula. The LPGE results of two-probe device based on group-V monolayer materials are shown in Section 3. Section 4 serves as the conclusion part.

\section{Computational details}
To study the optoelectronic properties of group-V materials, we first optimized the atomic structure of the monolayer As, Sb, and Bi (space group $Pmn2_1$, point group $ C_{2v}$), respectively. Here, they are relaxed when the residual force on each atom is smaller than 0.01 \textrm{eV {\AA}$^{-1}$} using the Vienna \emph{ab initio} simulation package (VASP)\cite{Kresse1993, Kresse1996}. 400 \textrm{eV} was chosen as the kinetic energy cutoff for the calculations. The plane wave basis set with the projector augmented wave (PAW)\cite{Bl1994} pseudo potential was adopted. The exchange-correlation potentials were approximated by the Local density approximation (LDA). To avoid the interaction between repeat images, a vacuum space of about 20 \textrm{{\AA}} perpendicular to the plane (direction of $y$ axis, as shown in Fig. 1) was used. In all geometry optimizations, van der Waals (vdW) correction (in the Grimme-D2 approach) was considered. The optimized buckling height (\textit{h}) are -0.17 \textrm{{\AA}}, -0.41 \textrm{{\AA}} and -0.59 \textrm{{\AA}} for monolayer As, Sb, and Bi, respectively (as shown in Fig. 1(a), $h=y_1-y_2$). The detailed parameters (such as lattice constants) are summarized in Table 1, which are in good agreement with previous calculations\cite{xiao2018, lu2015}. Since the LDA method always underestimates the band gap of semiconductors, the calculations of electronic properties are also conducted using the Heyd-Scuseria-Ernzerhof (HSE06) hybrid functional, which is constructed by mixing the PBE and Hartree-Fock (HF) functionals\cite{heyd2004, Krukau2006}. The screening parameter of HSE06 is 0.2 \textrm{{\AA}$^{-1}$}. The results show that the band structures calculated by the two methods only shift the bottom of the conduction band and the top of the valence band upward or downward, and their general characteristics are unchanged. Due to a large amount of calculation, we used the LDA method to analyze the optoelectronic properties of the two-probe system.

Starting from the above atomic structures obtained by minimizing total energy, we investigated the LPGE of the two-probe devices based on As, Sb, and Bi monolayers by carrying out non-equilibrium Green's function(NEGF) combined with density functional theory (DFT) formalism\cite{taylor2001, Taylor2001-2, waldron2006, zhai2014, lei2014, saraiva2016}, as implemented in the Nanodcal transport package\cite{nanodcal2, taylor2001}. In the subsequent NEGF-DFT numerical calculations, the linear combinations of atomic orbital (LCAO) basis sets and the standard norm-conserving nonlocal pseudo-potentials\cite{kleinman1982} were adopted; the Local density approximation was applied for the exchange-correlation potential; the wave functions and the other physical quantities were expanded with LCAO at the double-$\zeta$ polarization (DZP) level. The \textit{k}-point mesh is adopted as 13$\times$1$\times$13 for the self-consistent calculation of the lead region. In the photocurrent transport calculation of the two-probe device structures, 100$\times$1$\times$1 $k-$space grid was used in the armchair direction, 1$\times$1$\times$100 $k-$space grid was used in the zigzag direction. The self-consistency is deemed to be achieved when the total energy, hamiltonian, and density matrix are converged to an accuracy of 1$\times$10$^{-5} $a.u.

\begin{table}
\centering
{\small{\bf Table 1 } Lattice constants (a, b) and buckling height (\textit{h}) parameters of monolayer As, Sb, and Bi.\\
\vspace{2mm}
\begin{tabular}{cccc}
\hline
{Elements} & {a~({\AA})} & {b~({\AA})} & {\textit{h}~({\AA})}\\\hline
{As}       & {3.72}      & {4.23}      & {-0.17} \\
{Sb}       & {4.25}      & {4.43}      & {-0.41} \\
{Bi}       & {4.39}      & {4.57}      & {-0.59}\\
\hline
\end{tabular}}\label{table 1}
\end{table}

\begin{figure}%[!Htb]
\centering
\includegraphics[scale=0.5]{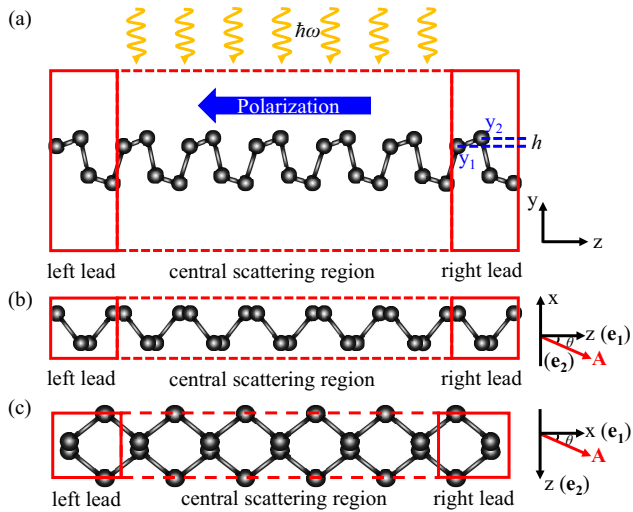}
\caption{Schematic plot of the two-probe device structure based on the group-V monolayer. (a) The side view of the relaxed configuration along the armchair direction; (b) The top view of the relaxed configuration along the armchair direction. Here, the length of central scattering region is 5*b. (c) The top view of the relaxed configuration along the zigzag direction. Here, the length of central scattering region is 4.5*a. It is divided into the left lead, right lead, and central scattering region where the light impinges with energy $\hbar \omega$. Here, $h=y_1-y_2$, the blue arrow represents the system's polarization direction. \textbf{A} is the electromagnetic vector potential inside the \textit{$x-z$} plane. ${\bf e_1}$ and ${\bf e_2}$ are two unit vectors for characterizing the polarization of the light. $\theta$ denotes the polarization angle for the linearly polarized light.}\label{fig1}
\end{figure}

In the two-probe structure (see Fig. 1), the photocurrent can be generated by the linearly polarized light shining on the central scattering region which is indicated by the box drawn with the red dashed line. In general, the direction of the current flow is from the lead to the central region. For simplicity, in the following, we consider the photocurrent flowing in the left lead of the two-probe structure. Firstly, we give the photocurrent $J_{L,s}^{(ph)}$, which is calculated in the first-order Born approximation with the following formula \cite{xie2015, lei2014, Chen2012},
\begin{eqnarray}
\label{photocurrent}
J^{(ph)}_{L,s}=\frac{ie}{h}\int{{\rm Tr}\{\Gamma_{L}[G^{<}_{ph}+f_L(E)(G^{>}_{ph}-G^{<}_{ph})]\}}_{ss}dE,
\end{eqnarray}
where $L$ indicates the left lead and $s$ indicates the spin component ($s$ = $\uparrow$, $\downarrow$); \textit{e} and \textit{h} are the electron charge and the Planck's constant; $\Gamma_L=i(\Sigma_L^r-\Sigma_L^a)$ is the linewidth function denoting the coupling between the central scattering region and the left lead, and $\Sigma^r_L=[\Sigma^a_L]^{\dagger}$ is the retarded self-energy due to the presence of the left lead; $f_L(E)$ is the Fermi-Dirac distribution function of the left lead; $G^{</>}_{ph}=G_0^{r}\Sigma_{ph}^{</>}G_0^{a}$ is the lesser/greater Green's function including electron-photon interaction\cite{henrickson2002}, where $\Sigma_{ph}^{</>}$ is the self-energy due to the presence of the electron-photon interaction, $G_0^{r/a}$ is the retarded/advanced Green's functions without photons. The information about polarization of the light is included in the self energy and can be characterized by a complex vector $\bf{e}$. For the linearly polarized light, $\bf{e}=\cos{\theta}\bf{e_1}+\sin{\theta}\bf{e_2}$, where $\theta$ is the angle formed by the polarized direction with respect to the vector $\bf{e_1}$ ($z$ axis when the transport direction is armchair direction; $x$ axis when the transport direction is zigzag direction). In our numerical calculations, the light is incident along the $-y$ direction (as shown in Fig. 1). For simplicity, we introduce a normalized photocurrent (photoresponse)\cite{henrickson2002, Chen2012}, which can be written as,
\begin{eqnarray}
\label{R}
R_s=\frac{J_{L,s}^{(ph)}}{eI_\omega},
\end{eqnarray}
where $J_{L,s}^{(ph)}$ is the photocurrent defined in Eq. (\ref{photocurrent}); $I_\omega$ is the photon flux defined as the number of photons per unit time per unit area. Note that the photoresponse has dimension of area, $a_0^2/photon$ where $a_0$ is the Bohr radius. The charge photocurrent ($I_c$) and spin photocurrent ($I_s$) can be defined as,
\begin{eqnarray}
\label{I_c}
I_c=R_\uparrow+R_\downarrow,\
I_s=R_\uparrow-R_\downarrow
\end{eqnarray}
where the $R_{\uparrow/\downarrow}$ represents the photoresponse with spin up or spin down component.
\section{Results and discussion}
Before analyzing the photoresponse characteristics of the two-probe device structure based on the group-V monolayer, we first discuss the influence of the spin-orbit coupling (SOC) effect from the band structure (BS) and the joint density of states (JDOS) of the bulk materials\cite{luo2020,Dresselhaus2001}, as shown in Fig. 2. Here, the JDOS can be written as: $J_{cv}(\hbar\omega)=\int\frac{2}{(2\pi)^2}\delta[E_c(\textbf{k})-E_v(\textbf{k})-\hbar\omega]d\textbf{k}$ when excited with a photon of frequency $\omega$. The $E_c(\textbf{k})$ and $E_v(\textbf{k})$ are the energies of electronic states at \textbf{k} point in the conduction and valence bands, respectively. Fig. 2(a-c) compares the band structures of As, Sb, and Bi, respectively, with and without SOC. It can be seen that SOC has rather little influence on the band structure of As, especially the band around the Fermi level, which is almost the same. However, for Sb and Bi, it is evident that SOC significantly influences their band structure, especially for Bi; the band-splitting effect near the Fermi level is distinguishable under SOC. This indicates that the impact of SOC can be ignored in calculating the photoresponse based on As monolayer. At the same time, the influence of SOC should be considered in analyzing photoresponse characteristics based on the Sb and Bi monolayer. Since the JDOS measures the number of allowed optical transitions from the valence band to the conduction band, Fig. 2(d-f) further investigates the JDOS distribution of As, Sb, and Bi with and without SOC. For As, the distribution of JDOS with and without SOC is almost the same, indicating that the SOC effect can be ignored in As monolayer. However, for Sb and Bi, the distribution of JDOS with and without SOC is different. Therefore, we will analyze the photoresponse characteristics of As monolayer without SOC and Sb and Bi monolayer with SOC.

\begin{figure*}%[!Htb]
\centering
\includegraphics[scale=0.8]{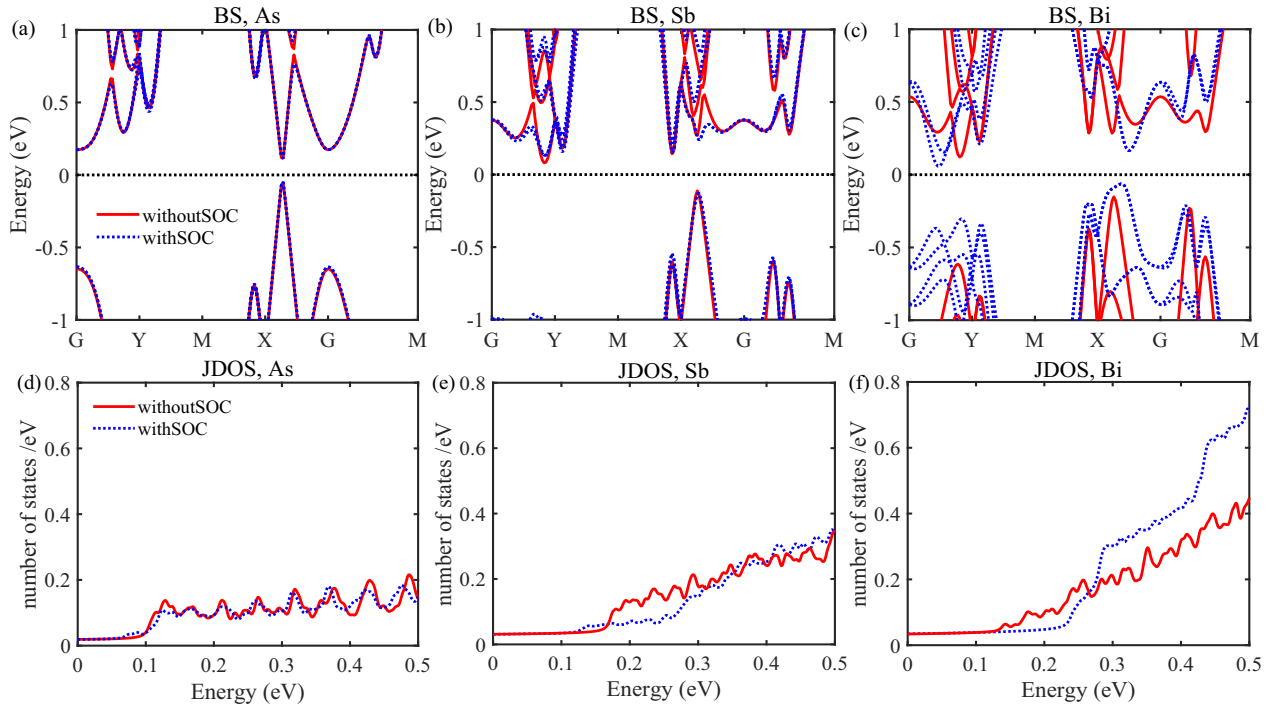}
\caption{(a-c) The band structures (BS) of As, Sb, and Bi with and without SOC. (d-f) The joint density of states (JDOS) of As, Sb, and Bi with and without SOC.}\label{fig2}
\end{figure*}

In the following, we present the LPGE of a two-probe device structure based on the group-V monolayer, as shown in Fig. 1. The whole system is divided into three regions, the central scattering region, and the left/right lead, which is indicated by the box drawn with a solid red line. And the leads extend to the electron reservoir into infinity. Here, we chose the linearly polarized light, which propagates along the $-y$ direction and irradiate the entire central scattering region of the device. Similar to the structure of black phosphorus, the primitive unit cell of monolayer As, Sb, Bi consists of two atomic layers with four atoms, two in the upper sublayer and the other two in the lower sublayer (as shown in Fig. 1(a)). The difference is that the black phosphorus structure has spatial inversion symmetry. In contrast, the symmetry of As, Sb, and Bi systems are broken due to the lattice distortion with out-of-plane atomic buckling, which causes electric charge accumulations at the outmost atoms, leading to spontaneous in-plane polarization along the $z$ axis (the blue arrow in Fig. 1(a) represent the direction of polarization of the system). Interestingly, since the group-V monolayer retains the mirror-symmetrical $M^x$, it produces the expected photoresponse characteristics, which we will introduce in detail later. Fig. 3(a, b) shows the photoresponse of As versus polarization angles $\theta=[0:15:180]$ at different photon energies in the armchair and zigzag direction. At first glance, the photoresponse in the armchair direction exhibits a cosine curve character, while the zigzag direction exhibits a sine curve character. In addition, we found that the magnitude and sign of photoresponse change for different photon energies, as shown in Fig. 3(a, b). This can be explained by the fact that the electrons are excited from the valence bands to the conduction bands under light irradiation. If the curvatures of the conduction band are different for +\textbf{k} and -\textbf{k}, the electrons excited to the conduction bands will have unbalanced motion, which results in the generation of a photocurrent. Depending on the photon energy, the electrons are activated to different \textit{k} points in conduction bands and obtain different band velocities. The summation of all activated electrons determines the sign of the photocurrent with other velocity distributions. So the character of the coefficient varies for different photon energies\cite{xie2015}. Fig. 3(c) shows the photoresponse of As versus the photon energies in the armchair and zigzag direction when $\theta = 0^\circ$. It can be seen that the photoresponse in the armchair direction is greater than that in the zigzag direction ($R_{ac}/R_{zz} \sim 10^5$), which presents giant anisotropy\cite{chu2018,ganichev2002-2,xie2015}. And the larger anisotropy in photoresponse can serve as an effective method in the determination of the lattice orientation of As monolayer.

\begin{figure*}%[!Htb]
\centering
\includegraphics[scale=0.8]{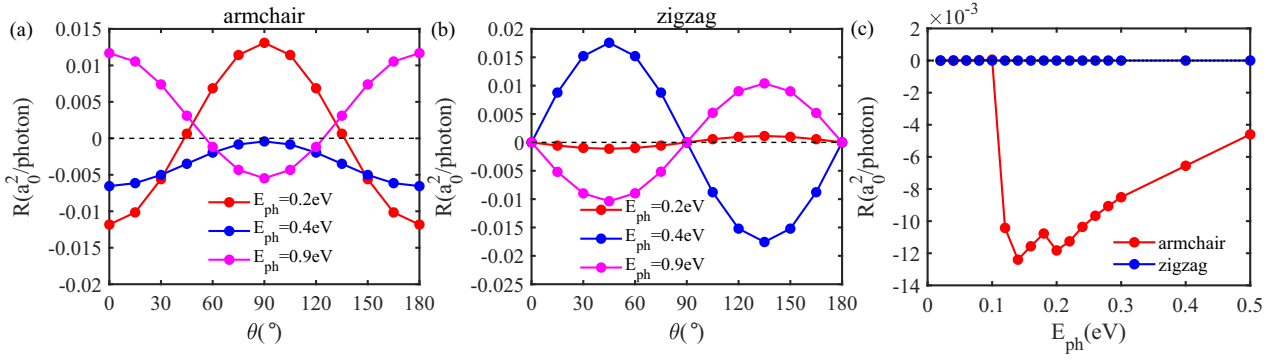}
\caption{The calculated photoresponse of As without SOC. (a) The photoresponse versus polarization angles $\theta$ at different photon energies in the armchair direction. (b) The photoresponse versus polarization angles $\theta$ at different photon energies in the zigzag direction. (c) The photoresponse versus the photon energies in the armchair (red circles solid line) and zigzag (blue circles solid line) direction when $\theta = 0^\circ$. The buckling height is -0.17 \textrm{{\AA}}.}\label{fig3}
\end{figure*}

\begin{figure}%[!Htb]
\centering
\includegraphics[scale=0.5]{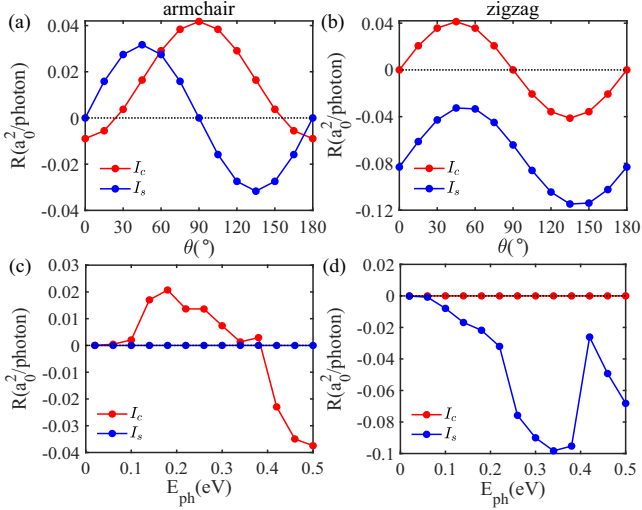}
\caption{The calculated photoresponse of Sb with SOC. (a, b) Charge photocurrent and spin photocurrent versus polarization angles $\theta$ in the armchair and zigzag direction, respectively, where photon energy is fixed as 0.4 eV. (c, d) Charge photocurrent and spin photocurrent versus photon energies in the armchair and zigzag direction, respectively, where $\theta$  = $0^ \circ$.}\label{fig4}
\end{figure}

For calculating the photoresponse of the two-probe device structure based on Sb and Bi monolayer, we included the SOC effect. We examined the charge and spin photocurrent along the armchair and zigzag directions, as shown in Fig. 4 and 5. We first analyzed the charge and spin photocurrent versus the light polarization angles $\theta$ for the Sb monolayer, as shown in Fig. 4(a, b). In the armchair direction, when the polarization angle $\theta = 0^\circ$, $90^\circ$, and $180^\circ$, there is no spin photocurrent generated. However, the corresponding charge photocurrent is nonzero. This means that only charge current can be produced in these situations. Otherwise, the spin-polarized photocurrent can be produced by tuning the polarization. In contrast, in the zigzag direction, the charge photocurrent is zero when $\theta = 0^\circ$, $90^\circ$, and $180^\circ$. Interestingly, the pure spin photocurrent with finite magnitude can be obtained at these polarization angles. To further demonstrate the dependence of photocurrent on photon energy, Fig. 4(c, d) analyze the charge and spin photocurrent versus photon energies when the polarization angle $\theta = 0^\circ$. As the photon energy increases, the spin photocurrent in the armchair direction is always zero, resulting in the generation of net charge current. Conversely, in the zigzag direction, there is no charge current, leading to the generation of pure spin current. The same physics can be acquired for the Bi monolayer, as shown in Fig. 5(a, b). The net charge current is generated in the armchair direction, and a pure spin current is obtained in the zigzag direction. Based on these results, let us refer back to the charge photocurrent of As monolayer in Fig. 3(c). Since the As monolayer's calculation does not considered the SOC effect, only the charge photocurrent was studied. From Fig. 3(c), we found that the charge current is always zero in the zigzag direction. However, the charge current can be finite in the armchair direction. Note that these results are consistent with that of Sb and Bi monolayer simulated with SOC.

\begin{figure*}%[!Htb]
\centering
\includegraphics[scale=0.8]{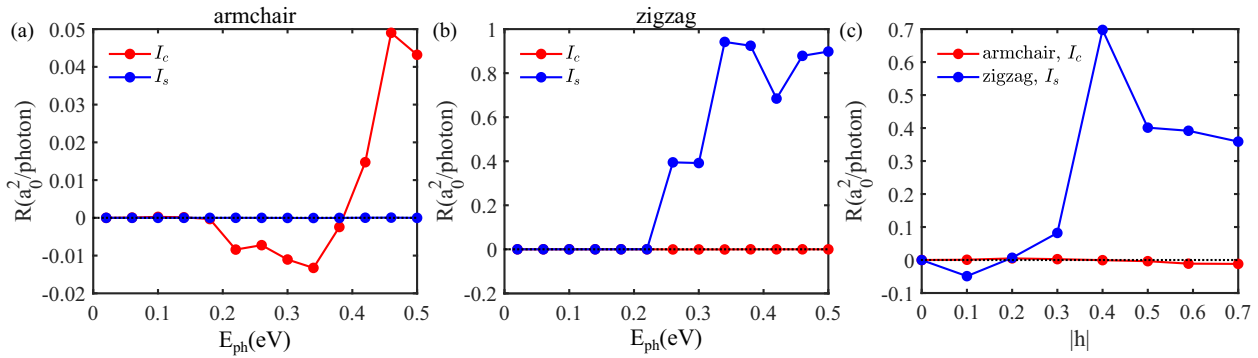}
\caption{The calculated photoresponse of Bi with SOC. (a, b) Charge and spin photocurrent versus photon energies in the armchair and zigzag direction, respectively, where $\theta$  = $0^ \circ$. (c) Charge current in the armchair direction and spin current in the zigzag direction versus the magnitude of buckling heights $|h|$, where $\theta$  = $0^ \circ$, $E_{ph} = 0.3$ eV.}\label{fig5}
\end{figure*}

To further alternatively understand the physics behind the generation of charge and spin photocurrent, we wish to start with the material's symmetry. It is known that ferroelectrics intrinsically exhibit the bulk photovoltaic (BPV) effect or bulk spin photovoltaic (BSPV) effect due to the fundamental requirement of spontaneous inversion symmetry breaking. It can generate charge current or spin current under light illumination. The nonlinear optical (NLO) charge current or spin current under light with frequency $\omega$ can be expressed as\cite{Xu2021}
\begin{eqnarray}
\label{NLO}
J^{a,s^i}=\sum_{\Omega=\pm\omega}\sigma^{a,s^i}_{bc}(0;\Omega,-\Omega)E^b(\Omega)E^c(-\Omega),
\end{eqnarray}
where $E(\omega)$ is the Fourier component of the electric field at angular frequency $\omega$. $\sigma^{a,s^i}_{bc}$ is the NLO conductivity and expressed as
\begin{eqnarray}
\label{conductivity}
\sigma^{a,s^i}_{bc}(0;\omega,-\omega)&=&-\frac{e^2}{\hbar^2\omega^2}\int\frac{d\textbf{k}}{(2\pi)^3}
\sum_{mnl}\frac{f_{lm}v^{b}_{lm}}{\omega_{ml}-\omega+i/\tau}
\nonumber \\
&*&(\frac{j^{a,s^i}_{mn}v^c_{nl}}{\omega_{mn}+i/\tau}-\frac{v^c_{mn}j^{a,s^i}_{nl}}{\omega_{nl}+i/\tau}),
\end{eqnarray}
where $a$ indicates the direction of the current, while $b$ and $c$ are the polarization direction of polarized light. $s^i$ with $i = x, y, z$ is the spin polarization, while $s^0$ represents charge current. $j^{a,s^i}=1/2(v^as^i+s^iv^a)$ indicate the spin current ($i\neq0$) and charge current ($s^0=e$). The numerators in Eq. (5) are composed of terms with the format $N_{mnl}^{iabc}=j_{mn}^{a,s^i}v_{nl}^bv_{lm}^c (i\neq0)$ for spin current and $N_{mnl}^{0abc}=v_{mn}^av_{nl}^bv_{lm}^c (i=0)$ for charge current. For the monolayer of the group-V we studied, since it has mirror symmetry $M^x$, the following relationship can be obtained
\begin{eqnarray}
M^xv^a_{mn}(\textbf{k})=(-1)^{\delta_{xa}}v^a_{mn}(\bf{k'})
\end{eqnarray}
and
\begin{eqnarray}
M^xs^i_{mn}(\textbf{k})=-(-1)^{\delta_{xi}}s^i_{mn}(\bf{k'})
\end{eqnarray}
When the mirror-symmetrical $M^x$ is applied to the conductivity $\sigma^{a,s^i}_{bc}$, it can be found that when the polarization direction of the polarized light is along the $x$ axis direction, $\sigma^{x,s^0}_{x,x} = 0,\ \sigma^{z,s^0}_{x,x} \neq 0,\ \sigma^{x,s^i}_{x,x} \neq 0,\ \sigma^{z,s^i}_{x,x} = 0$, that is, a non-zero charge current and a non-zero spin current are generated in the $z$ direction and $x$ direction, respectively. Similarly, when the polarization direction of the polarized light is along the $z$ axis direction, $\sigma^{x,s^0}_{z,z} = 0,\ \sigma^{z,s^0}_{z,z} \neq 0,\ \sigma^{x,s^i}_{z,z} \neq 0,\ \sigma^{z,s^i}_{z,z} = 0$, i.e., the $x$ direction produces a non-zero spin current, while the $z$ direction produces a non-zero charge current. For the detailed calculation process please refer to ref. 58. Therefore, in the group-V monolayer, we studied when the polarization direction of linearly polarized light is parallel or perpendicular to the transport direction of the system ($\theta$ = $0^ \circ$ or $90^ \circ$), a net charge current can be generated in the armchair direction. In addition, a pure spin current can be obtained in the zigzag direction, consistent with our numerical results above.

Last but not least, since the degree of buckling $h$ can be effectively tuned by charge doping\cite{lu2015}, it will be essential to know if the photocurrent generated by PGE can be further adjusted. Moreover, we calculated the charge photocurrent in the armchair direction and spin photocurrent in the zigzag direction of Bi monolayer versus the magnitude of buckling heights, as shown in Fig. 5(c). Obviously, the charge current in the armchair direction and the spin current in the zigzag direction oscillate with the increase of the buckling height $|h|$. This is because the shift current is the main contribution to the linearly photogalvanic effect (LPGE) generated photocurrent, the shift current depends not only on the buckling height, but also on the density of states, velocity matrix elements, and shift-vector matrix elements\cite{Fregoso2017}. From response theory, it is found that the short-circuit current on a device is proportional to the sum of polarization differences\cite{Fregoso2017}. This indicates that the larger polarization of the ferroelectric material, the larger photocurrent can be obtained, which suggests that 2D ferroelectrics are potential candidates for photovoltaic materials.

\section{Conclusion}
To summarize, we investigated the linear photogalvanic effect of two-probe devices based on group-V (As, Sb, and Bi) monolayer from first-principles calculations. By calculating the photoresponse of As in the armchair and zigzag direction, it is found that there is apparent anisotropy. More interestingly, in the two-probe device based on Sb and Bi monolayer, when the polarization direction of linearly polarized light is parallel or perpendicular to the transport direction, net charge current and pure spin current can be generated in the armchair and zigzag direction, respectively. This is also understood by analyzing the nonlinear optical coefficient. More importantly, it is also found that the net charge current generated in the armchair direction and the pure spin current generated in the zigzag direction can be further tuned with the increase of the material's buckling height $|h|$, which paves the novel applications of group-V monolayers in optoelectronics and opto-spintronics.

\section*{Acknowledgements}
We gratefully acknowledge the the support from the National Key R\&D Program of China under Grant No. 2022YFA1404003, the National Natural Science Foundation of China (Grants No. 12074230, 12174231), the Natural Science Foundation of Shanxi Normal University, the Fund for Shanxi ``1331 Project", and Shanxi Province 100-Plan Talent Program, Fundamental Research Program of Shanxi Province through 202103021222001, 202203021212397. This research was conducted using the High Performance Computer of Shanxi University.

\bigskip

\noindent{$^{\dagger)}$ Electronic address: zhanglei@sxu.edu.cn}

%\addcontentsline{toc}{chapter}{References}

%\bibstyle{}
%\bibliography{ref} %You need to replace "rsc" on this line with the name of your .bib file
%\end{CJK*}
\end{document}